\documentstyle[art10,epsfig]{article}

\newcommand\beq{\begin{eqnarray}}
\newcommand\eeq{\end{eqnarray}}

\newcommand\la{\langle}
\newcommand\ra{\rangle}
\newcommand\rar{\rightarrow}

\newcommand\q{\la\bar{q}q\ra}

\newcommand\s{\la\bar{s}s\ra}

\newcommand\ul{\underline}

\begin{document}

\title{Multiquark states and QCD sum rules}
\author{
Seungho Choe\thanks{E-mail: schoe@phya.yonsei.ac.kr} \\
{\it Department of Physics, Yonsei University}\\
{\it Seoul 120--749, Korea}}

\date{}
\maketitle

\begin{abstract}
There have been arguments about hadronic molecules, which are weakly-bound
states of two or more hadrons. We investigate the possibility of some
candidates ($f_0$(980), $a_0$(980), $f_0$(1500), $f_0$(1710), etc.)
using QCD sum rule approach and compare our results with multiquark states
in the MIT bag model.
We find that
$f_0$(1500), $f_0$(1710) can be good candidates for vector-vector
molecule-type multiquark states.
\end{abstract}

\section{Introduction}

One can classify exotics by the number of quarks plus antiquarks they contain;
i.e., glueballs, hybrid mesons, hybrid baryons, four quark states, and so on.
Four quark states have two quarks and two antiquarks. A special case which
we discuss below are hadronic molecules
[($q\bar{q}$)($q\bar{q}$)].\footnote{For a recent review of
the hadronic molecules,
see Ref.\cite{Barnes94}.}

The first serious estimation of the interaction between quarks in the four
quark system was done within the MIT bag model by Jaffe\cite{Jaffe77}.
He found
that the scalar states $O^{++}$ have lowest mass, and interpreted the
$f_0$ (980) and $a_0$ (980) as four quark states.
Table \ref{table_bag} shows several
states of scalar mesons and their masses which are discussed
in our study.
For the details, see Ref.\cite{Jaffe77}.
\begin{table}[t]
\caption{The predicted $Q^2 \bar{Q}^2 ~0^+$ mesons. Masses are
quoted to the nearest 50 MeV.}
\label{table_bag}
\begin{center}
\begin{tabular}{clc}
\hline
     SU(3) multiplet &    State    &   Mass (MeV) \\
\hline
 $\ul{9}$ &   $ C^0 (\ul{9}) = \frac{\sqrt{3}}{2} \pi \pi
                     + \frac{1}{2} \eta_0 \eta_0$        & 650      \\
	&   $ C^s (\ul{9}) = \frac{1}{\sqrt{2}} K \bar{K} +
                  \frac{1}{\sqrt{2}} \eta_0 \eta_s$      & 1100     \\
	&   $ C^s_\pi(\ul{9}) = - \frac{1}{\sqrt{2}} K \bar{K}
                        - \frac{1}{\sqrt{2}} \eta_s \pi$  & 1100 \\
\hline
   $\ul{36}$ & $C^0 (\ul{36}) = -\frac{1}{2} \pi \pi
                    + \frac{\sqrt{3}}{2} \eta_0 \eta_0$    &  1150    \\
	     &	 $ C^s (\ul{36}) = \frac{1}{\sqrt{2}} K \bar{K} -
                  \frac{1}{\sqrt{2}} \eta_0 \eta_s$       &  1550   \\
	     &	$ C^s_\pi(\ul{36}) =  \frac{1}{\sqrt{2}} K \bar{K}
                        - \frac{1}{\sqrt{2}} \eta_s \pi$   &  1550 \\
\hline
    $\ul{9}^*$ &  $ C^0 (\ul{9^*}) = \frac{\sqrt{3}}{2} \rho \rho
                         + \frac{1}{2} \omega \omega$     &  1450   \\
	       & $ C^s (\ul{9^*}) = \frac{1}{\sqrt{2}} K^* \bar{K}^{*}
                     + \frac{1}{\sqrt{2}} \omega \phi$    & 1800  \\
          &  $C^s_\pi (\ul{9^*}) = - \frac{1}{\sqrt{2}} K^* \bar{K}^{*}
                       - \frac{1}{\sqrt{2}} \rho \phi$    & 1800  \\
\hline
    $\ul{36}^* $ & $ C^0 (\ul{36^*}) = -\frac{1}{2} \rho \rho
                        + \frac{\sqrt{3}}{2} \omega \omega$ & 1800  \\
	  &    $C^s (\ul{36^*}) = \frac{1}{\sqrt{2}} K^* \bar{K}^{*}
                     - \frac{1}{\sqrt{2}} \omega \phi$    &  2100   \\
	 &  $C^s_\pi (\ul{36^*}) =  \frac{1}{\sqrt{2}} K^* \bar{K}^{*}
                       - \frac{1}{\sqrt{2}} \rho \phi$   & 2100 \\
\hline
\end{tabular}
\end{center}
\end{table}
The bag model results
were confirmed in a Nonrelativistic Quark Model calculation
by Weinstein and Isgur\cite{WI82}. These authors found that the predominant
component in the $f_0$ and the $a_0$ wave functions was $K\bar{K}$, as a
mesonic molecule.
Very interesting proposals were made separately by
T\"ornqvist\cite{Tornqvist91} and by Dooley {\it et al.}\cite{DSB92},
in suggesting the existence of
vector meson molecules.
There are several signatures for
 hadronic molecules\cite{Barnes94}.

In this paper we investigate the possibility of hadronic molecules for
several candidates using QCD sum rule approach\cite{SVZ79,RRY85}.
They are $f_0$(980), $a_0$(980), $f_0$(1500), and
$f_0$(1710)\footnote{The spin state of $f_J$(1710) is not
clarified at present\cite{PRD}.}.
$f_0$(980) and $a_0$(980) are candidates of $K\bar{K}$ molecule, and
$f_0$(1500) is a $\rho\rho$ or $\rho\rho + \omega\omega$ molecular state.
$f_0$(1710) is a $K^* \bar{K}^*$ or $K^* \bar{K}^* + \omega\phi$
molecular state.
Of course there are another interpretations on
$f_0(1500)$ and $f_0(1710)$\cite{Weingarten94,AC96,Genovese96};
i.e. glueball states
 or mixed states of a $\bar{s}s$ meson and a digluonium, etc.
We predict masses of these particles with appropriate interpolating
fields for molecular states
(e.g., ``molecular-like" interpolating field
($\bar{q}\Gamma q)(\bar{q}\Gamma q)$).
In addition to this, we assume that the four quark states
in Jaffe's notations as hadronic molecules
and then calculate their masses, and compare our results with Jaffe's.

\section{$K\bar{K}$ Molecule}

\begin{figure}[t]
\begin{center}
\leavevmode
\epsfig{file=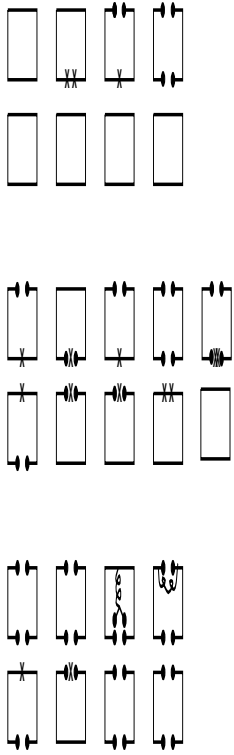,height=3cm,width=10cm,angle=-90}
\end{center}

\caption{Diagrams of 2 loop-type. Solid lines are the quark propagators and
curly line represents the gluon propagator.
Dots denotes the quark condensates
and cross represents the mass correction from
the strange quark. }
\label{2loop}
\end{figure}

\begin{figure}
\begin{center}
\leavevmode
\epsfig{file=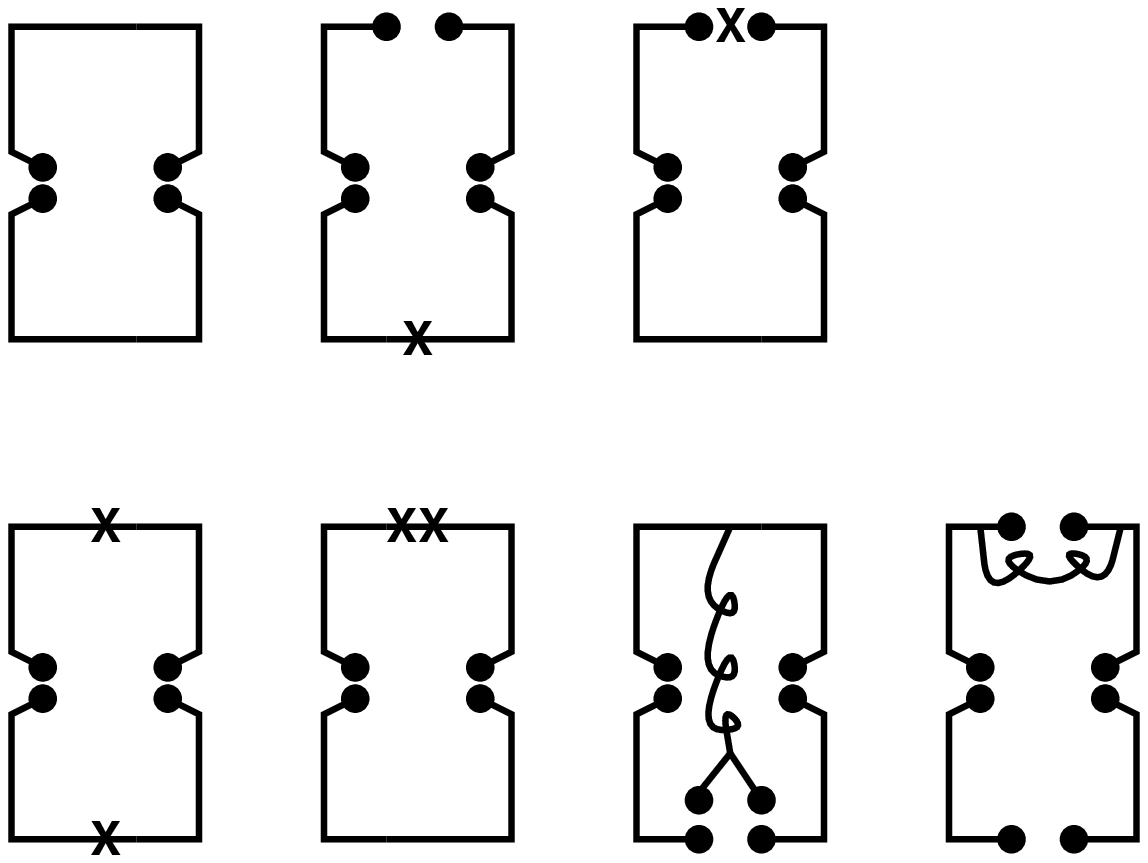,height=2.5cm,width=7.8cm,angle=-90}

\end{center}
\caption{Diagrams of 1 loop-type. Solid lines are the quark propagators and
curly line represents the gluon propagator.
Dots denotes the quark condensates
and cross represents the mass correction from
the strange quark. }
\label{1loop}
\end{figure}

Let's consider the following correlator:
\beq
\Pi (q^2) = i \int d^4x e^{iqx}\langle T ( J(x) J^\dagger(0) )\rangle ,
\eeq
where $J(x) = (\bar{u}(x) i\gamma^5 s(x))(\bar{s}(x) i\gamma^5 u(x))
 + (\bar{d}(x) i\gamma^5 s(x))(\bar{s}(x) i\gamma^5 d(x))$ corresponds
 to the multiquark interpolating field for $f_0$(980) state.
This is the
$K^0 \bar{K}^{0} + K^+ K^-$
state (isospin I=0).
Then, in the OPE side we get
\beq
\Pi_{OPE} (q^2) &=& - \frac{1}{\pi^6 2^{14} 5} q^8 ln(-q^2)
	      + \frac{m_s^2}{\pi^6 2^{11}} q^6 ln(-q^2)
\nonumber\\
&+& \frac{m_s}{\pi^4 2^8} (2 \q - \s ) q^4 ln (-q^2)
-\frac{1}{\pi^2 2^7 3}(24\q \s
		+ 2\q ^2 + \s ^2) q^2 ln(-q^2)
\nonumber\\
&-& \frac{m_s^2}{\pi^2 2^6}(2\q^2 - 8\q \s
		      + \s^2) ln(-q^2)
+\frac{m_s}{2^4 3}(6\q^2 \s
	     - 4\q \s^2) \frac{1}{q^2}
\nonumber\\
&+& \frac{\pi\alpha_s}{3^5}(102\q^2 \s^2
	     - 16\q^3 \s
             - 20\q \s^3) \frac{1}{q^4} ,
\label{ope_f0}
\eeq
where $m_s$ is a strange quark mass.
On the other hand, for $a_0$(980)
we take the interpolating field as
$K^0 \bar{K}^{0} - K^+ K^-$
state (isospin I=1).
Then, we have
\beq
\Pi_{OPE} (q^2) &=& - \frac{1}{\pi^6 2^{14} 5} q^8 ln(-q^2)
	      + \frac{m_s^2}{\pi^6 2^{11}} q^6 ln(-q^2)
\nonumber\\
&+& \frac{m_s}{\pi^4 2^8} (2 \q - \s ) q^4 ln (-q^2)
-\frac{1}{\pi^2 2^7 3}(24\q \s
		+ \s ^2) q^2 ln(-q^2)
\nonumber\\
&-& \frac{m_s^2}{\pi^2 2^6}(4\q^2 - 8\q \s
		      + \s^2) ln(-q^2)
+\frac{m_s}{2^4 3}(8\q^2 \s
	     - 4\q \s^2) \frac{1}{q^2}
\nonumber\\
&+& \frac{\pi\alpha_s}{3^5}(178\q^2 \s^2
	     - 24\q^3 \s
             - 20\q \s^3) \frac{1}{q^4} .
\label{ope_a0}
\eeq
In the above calculations we use two diagrams: Fig. \ref{2loop} and
Fig. \ref{1loop}.
Here, we neglect the contribution of gluon condensates
and concentrate on tree diagrams. We assume the vacuum saturation hypothesis
to calculate quark condensates of higher dimensions.
Similar calculation is found in Kodama {\it et al.}'s H-dibaryon sum rules
\cite{KOH94}.
%

In Eqs.(\ref{ope_f0}), (\ref{ope_a0}) above
the OPE sides have the following form:
\beq
\Pi_{OPE}(q^2) &=& a ~q^8 ln(-q^2) + b ~q^6 ln(-q^2) + c ~q^4 ln(-q^2)
+ d ~q^2 ln(-q^2)
\nonumber \\
&+& e ~ln(-q^2) + f ~\frac{1}{q^2} + g ~\frac{1}{q^4} ,
\eeq
where $a, b, c, \cdots, g$ are constants. Then we parameterize the
phenomenological side as
\beq
\frac{1}{\pi} Im \Pi_{phen} (s) = \lambda^2 \delta(m^2-s) +
             [-a~s^4 - b~s^3 - c~s^2 - d~s - e] \theta(s~-~s_0) ,
\eeq
where $s_0$ is a continuum threshold.
After Borel transformation we obtain a mass of $f_0$ and $a_0$ respectively.
The mass m is given by
\beq
m^2 &=& M^2 \times
\nonumber\\
&\{&-120a [1-e^{-s_0/M^2}(1+\frac{s_0}{M^2}+\frac{s_0^2}{2M^4}+
    \frac{s_0^3}{6M^6}+\frac{s_0^4}{24M^8}+\frac{s_0^5}{120M^{10}})]
\nonumber\\
&-& \frac{24b}{M^2} [1-e^{-s_0/M^2}(1+\frac{s_0}{M^2}+\frac{s_0^2}{2M^4}
		       +\frac{s_0^3}{6M^6}+\frac{s_0^4}{24M^8})]
\nonumber\\
&-& \frac{6c}{M^4} [1-e^{-s_0/M^2}(1+\frac{s_0}{M^2}+\frac{s_0^2}{2M^4}
		       +\frac{s_0^3}{6M^6})]
\nonumber\\
&-& \frac{2d}{M^6} [1-e^{-s_0/M^2}(1+\frac{s_0}{M^2}+\frac{s_0^2}{2M^4})]
\nonumber\\
&-& \frac{e}{M^8} [1-e^{-s_0/M^2}(1+\frac{s_0}{M^2})] - \frac{g}{M^{12}}~\}
~/
\nonumber\\
&\{&-24a [1-e^{-s_0/M^2}(1+\frac{s_0}{M^2}+\frac{s_0^2}{2M^4}+
      \frac{s_0^3}{6M^6}+\frac{s_0^4}{24M^8})]
\nonumber\\
&-& \frac{6b}{M^2} [1-e^{-s_0/M^2}(1+\frac{s_0}{M^2}+\frac{s_0^2}{2M^4}
		       +\frac{s_0^3}{6M^6})]
\nonumber\\
&-& \frac{2c}{M^4} [1-e^{-s_0/M^2}(1+\frac{s_0}{M^2}+\frac{s_0^2}{2M^4})]
\nonumber\\
&-& \frac{d}{M^6} [1-e^{-s_0/M^2}(1+\frac{s_0}{M^2})]
\nonumber\\
&-& \frac{e}{M^8} [1-e^{-s_0/M^2}] - \frac{f}{M^{10}} + \frac{g}{M^{12}}\},
\eeq
where we take $\q$ = --(0.230 GeV)$^3$, $\s=0.8\q$,
 $\alpha_s$ = 0.5,
and $m_s$ = 0.150 GeV throughout
this paper.
The continuum contribution is large, so this formula
has large uncertainties. We can not find a plateau for
the mass of $f_0$ and $a_0$.
Thus, we have to change our strategy.
Let's consider Fig. \ref{2loop} and Fig. \ref{1loop} again.
The diagrams in Fig. \ref{2loop} (hereafter we call it 2 loop-type)
are proportional to $N_c^2$ (where $N_c$ is a number of color) and
the diagrams in Fig. \ref{1loop} (hereafter 1 loop-type)
$N_c$. Hence,
the 1 loop-type diagrams are $1/N_c$ corrections to the 2 loop-type diagrams.
It means that the OPE side can be written as
\beq
\Pi_{OPE} (q^2) = N_c^2 ( 1\times {\rm ~2 ~loop-type}
                         + {1 \over N_c} \times {\rm ~1 ~loop-type} ) .
\eeq
According to Witten's arguments on large $N_c$ dynamics
\cite{Witten79},
there are no exotics in the leading order of $N_c$.
2 loop-type corresponds to the leading order, and it means that
the two kaons are flying without any interaction among themselves.
Therefore, our new strategy are as follows:
First, consider 2 loop-type only and vary the continuum threshold $s_0$ and
Borel interval $M^2$
in order that the mass should be 990 MeV
(the sum of two free kaon masses).
The Borel interval $M^2$ is restricted by the following conditions as usual:
OPE convergence and pole dominance.
Second, consider all diagrams (2 loop-type + 1 loop-type) and
get a new mass $m^{\prime}$ with the same $s_0$ and Borel interval
$M^2$ which are obtained
from the first step.
Third, compare  $m^\prime$ with 990 MeV. If $m^{\prime}$ is less than
990 MeV, it can be one signature for molecular-like multiquark states.
%
\begin{table}[t]
\caption{$K\bar{K}$ molecule}
\label{table_kk}
\begin{center}
\begin{tabular}{ c c c c c}
\hline
     &            & $s_0$ (GeV$^2$)  & $M^2$ (GeV$^2$)   & mass (GeV) \\
\hline
 $K\bar{K}$ & $N_c\rar \infty$
                           & 2.06   & 0.81 -- 0.98   & 0.990 \\
           & $K^0\bar{K}^{0} + K^+K^-$ &         &   & 1.031 \\
           & $K^0\bar{K}^{0} - K^+K^-$ &         &   & 1.000 \\
           & $K^+ K^+$                 &         &   & 0.968 \\
\hline
\end{tabular}
\end{center}
\end{table}

Our results are in Table \ref{table_kk}.
In the Table, one can see that
the masses of $f_0$ and $a_0$ are greater than the two kaons' mass (990 MeV).
Even the mass of $K^+ K^+$ state
is less than 990 MeV.
It is worthy to note that our results do not change
even though we take another value from
the case of ``$N_c \rar \infty$".
The masses of $f_0$ and $a_0$ are always greater than
the threshold, and $K^+ K^+$ state is always lower than the threshold.

\section{Vector-Vector Molecule}

In this section we consider vector-vector molecules, such as
$\rho\rho$ and $K^* \bar{K}^*$ molecules.
For the $\rho\rho$ we take $\rho^+ \rho^- + \rho^0 \rho^0$
state($f_0$(1500)).
There are two types in $K^* \bar{K}^*$ molecules:
isospin I=0 and I=1 states. These are
$K^{*0} \bar{K}^{*0} + K^{*+} \bar{K}^{*-}$ (I=0, i.e., $f_0$(1710)) and
$K^{*0} \bar{K}^{*0} - K^{*+} \bar{K}^{*-}$ (I=1).
We use the same procedure as in the case of $K\bar{K}$ molecule.
The results are in Table \ref{table_vector}.
%
\begin{table}[t]
\caption{$\rho\rho$ and $K^* \bar{K}^*$ molecule}
\label{table_vector}
\begin{center}
\begin{tabular}{c c c c c }
\hline
     &            & $s_0$ (GeV$^2$)  & $M^2$ (GeV$^2$)   & mass (GeV) \\
\hline
$\rho\rho$ & $N_c\rar \infty $ & 4.12 & 1.20 -- 2.00  & 1.540 \\
 & $\rho^0 \rho^0 + \rho^+ \rho^-$   &         &           & 1.499 \\
        & $\rho^+ \rho^+ $           &         &           & 1.542 \\
\hline
$K^* \bar{K}^*$ & $N_c\rar \infty$  & 5.36 & 1.25 -- 2.85 & 1.784  \\
        & $K^{*0}\bar{K}^{*0}+ K^{*+}K^{*-}$ & &           & 1.748 \\
        & $K^{*0}\bar{K}^{*0}- K^{*+}K^{*-}$ & &           & 1.753 \\
        & $K^{*+}K^{*+}$                      & &               & 1.791 \\
\hline
\end{tabular}
\end{center}
\end{table}

We see that there is a binding
and $f_0$ (1500) is a good candidate of vector-vector
molecular-like multiquark state.
For the case of $K^* \bar{K}^*$, two states (I=0 and I=1)
are lower than the threshold 1.784 GeV.
In this case $f_0$ (1710) is also a good candidate
of vector-vector molecular-like multiquark state.
In addition, as can be seen in the Table,
I=1 state ($K^{*0} \bar{K}^{*0} - K^{*+} \bar{K}^{*-}$)
can be another good candidate of vector-vector
molecular-like multiquark state.

\section{Discussion}

\begin{table}[t]
\caption{$K \bar{K}$ and $\pi \pi$ with Jaffe's notations
(($\cdots$) means Jaffe's result.)}
\label{table_kk_jaffe}
\begin{center}
\begin{tabular}{ c c c c c }
\hline
     &         & $s_0$ (GeV$^2$)  & $M^2$ (GeV$^2$)   & mass (GeV) \\
\hline
  $K \bar{K}$& $N_c\rar \infty $ & 2.08 &  0.79 -- 0.98  & 0.990 \\
  & $ \frac{1}{\sqrt{2}} K \bar{K} + \frac{1}{\sqrt{2}} \eta_0 \eta_s$
                                             &     &     & 0.995 (1100)\\
  & $- \frac{1}{\sqrt{2}} K \bar{K} - \frac{1}{\sqrt{2}} \eta_s \pi$
                                             &     &     & 0.973 (1100) \\
\cline{2-5}
     & $ \frac{1}{\sqrt{2}} K \bar{K} - \frac{1}{\sqrt{2}} \eta_0 \eta_s$
                                             &     &    & 1.021 (1550) \\
     & $ \frac{1}{\sqrt{2}} K \bar{K} - \frac{1}{\sqrt{2}} \eta_s \pi$
                                             &     &    & 1.008 (1550) \\
\hline
 $\pi \pi$ & $N_c\rar \infty $ & 2.08 &  0.79 -- 0.98  & 1.000 \\
     & $ \frac{\sqrt{3}}{2} \pi \pi + \frac{1}{2} \eta_0 \eta_0$
                                      &       &        & 1.039 (650) \\
\cline{2-5}
     & $ -\frac{1}{2} \pi \pi + \frac{\sqrt{3}}{2} \eta_0 \eta_0$
                                      &       &        & 1.001 (1150) \\
\hline
\end{tabular}
\end{center}
\end{table}

\begin{table}[t]
\caption{$K^* \bar{K}^*$ and $\rho \rho$ with Jaffe's notations
	 (($\cdots$) means Jaffe's result.)}
\label{table_vector_jaffe}
\begin{center}
\begin{tabular}{c c c c c }
\hline
       &         & $s_0$ (GeV$^2$)  & $M^2$ (GeV$^2$)   & mass (GeV) \\
\hline
$K^* \bar{K}^*$& $N_c\rar \infty $
                      & 5.37 &  1.24 -- 2.83  & 1.784 \\
  & $ \frac{1}{\sqrt{2}} K^* \bar{K}^{*} + \frac{1}{\sqrt{2}} \omega \phi$
                                             &     &   & 1.775 (1800) \\
  & $- \frac{1}{\sqrt{2}} K^* \bar{K}^{*} - \frac{1}{\sqrt{2}} \rho \phi$
                                             &     &   & 1.781 (1800)  \\
\cline{2-5}
  & $ \frac{1}{\sqrt{2}} K^* \bar{K}^{*} - \frac{1}{\sqrt{2}} \omega \phi$
                                             &     &   & 1.755 (2100)\\
  & $ \frac{1}{\sqrt{2}} K^* \bar{K}^{*} - \frac{1}{\sqrt{2}} \rho \phi$
                                             &     &   & 1.759 (2100) \\
\hline
$\rho \rho$ & $N_c\rar \infty $
               & 4.13 &  1.16 -- 2.03  & 1.540 \\
  & $ \frac{\sqrt{3}}{2} \rho \rho + \frac{1}{2} \omega \omega$
                                          &       &   & 1.507 (1450) \\
\cline{2-5}
  & $ -\frac{1}{2} \rho \rho + \frac{\sqrt{3}}{2} \omega \omega$
                                          &       &   & 1.538 (1800)\\
\hline
\end{tabular}
\end{center}
\end{table}

In Table \ref{table_kk_jaffe} and Table \ref{table_vector_jaffe}
we compare our results with that of Jaffe.
Detail calculations are given in Ref.\cite{Choe96}.
First, consider Table \ref{table_kk_jaffe}.
We present two cases, $K\bar{K}$ and $\pi\pi$ molecular states.
In the case of $\pi\pi$ state
when $N_c \rar \infty$, we can not set the mass
to that of the sum of two pion masses ($\sim$ 280 MeV).
So we take the same $s_0$ and $M^2$ as
those from the case of $K\bar{K}$, and
compare a magnitude; i.e. whether the mass is below or above the threshold.
Our results are much different from those of Jaffe.
We can not predict mass splittings as in the case of bag model.
Next, move on to Table \ref{table_vector_jaffe}.
The result of two cases ($K^* \bar{K}^*$ and $\rho \rho$ molecular states)
are presented.
We can not obtain the mass splitting as that from the bag model.
However, the masses of
$ \frac{1}{\sqrt{2}} K^* \bar{K}^{*} + \frac{1}{\sqrt{2}} \omega \phi$
($f_0$ (1710)) and
 $ \frac{\sqrt{3}}{2} \rho \rho + \frac{1}{2} \omega \omega$
 ($f_0$ (1500))
are lower than their respective threshold values. Thus,
we can think these as vector-vector
molecular-like multiquark state.\footnote{We also checked that $f_0(1500)$ with
the other normalization, i.e.
 $ {1 \over \sqrt{2}} \rho \rho + {1 \over \sqrt{2}} \omega \omega$,
 has a lower mass than the threshold.}
 Besides,
 $- \frac{1}{\sqrt{2}} K^* \bar{K}^{*} - \frac{1}{\sqrt{2}} \rho \phi$
 state may be another candidate of vector-vector molecular-like multiquark
 state which was proposed in \cite{DSB92}.
%
\begin{table}[t]
\caption{$\Lambda (1405)$  (tentative)}
\label{table_lambda}
\begin{center}
\begin{tabular}{c c c c c c }
\hline
   &       &         & $s_0$ (GeV$^2$)  & $M^2$ (GeV$^2$)   & mass (GeV) \\
\hline
       & $\sim$ $\rlap{/}{q}$ & $N_c\rar \infty$
                           & 3.553   & 1.4 -- 2.0   & 1.435 \\
       &              &    &         &              & 1.409 \\
\hline
\end{tabular}
\end{center}
\end{table}


In Table \ref{table_lambda} we present a result for the case of
$\Lambda (1405)$.
It has long been considered a candidate of $\bar{K}N$
bound state\cite{VJTR85,Jennings86}, since it
is just below the $\bar{K}N$ threshold.
We assume this a $\bar{K}-N$ molecular state,
and investigate the possibility of
hadronic molecule using the previous approach.
Because of its dimension the OPE side has two structures:
\beq
\Pi_{OPE} (q^2) = \Pi_1 (q^2) {\bf 1} + \Pi_q (q^2) {\bf \rlap{/}{q}} .
\eeq
We can obtain a mass from
$\Pi_1 (q^2)$ and $\Pi_q (q^2)$ respectively .
In the Table a result from $\Pi_q (q^2)$ is given.
In this case there is a binding also.
Note that this is a tentative result.

In summary,
we showed that $f_0$(1500), $f_0$(1710) are good candidates for vector-vector
molecular-like multiquark state.
As a possible modification to the cases of $f_0$(980), $a_0$(980)
we can consider the direct instanton effect\cite{Shuryak} in our calculation.
However, our work is still insufficient
to predict whether it is a molecular state or a molecule-type
multiquark state or even an ordinary quark-antiquark state
with large virtual quark-antiquark pair. So we have to
calculate another quantity, such as $\gamma\gamma$ decay widths
of these scalar mesons, and compare with
the experiments.
Recently, Close {\it et al.}\cite{CIK93}
suggested a related test for the $f_0$(980)
and $a_0$(980) involving the radiative decays $\phi \rar \gamma(f_0, a_0)$,
 which may be possible at DA$\Phi$NE and CEBAF.
Our methods may be applied to hadron-hadron scattering
because the study of molecules is a subtopic of the problem of determining
2 $\rar$ 2 hadron-hadron scattering amplitudes near threshold\cite{Barnes94}.

\bigskip

{\bf Acknowledgements}
\\

This work was done in collaboration with Prof. Su H. Lee.
It was supported in part by KOSEF through CTP
at Seoul National University.
The author thanks Prof. Y. Koike for his
hospitality during this beautiful workshop.


\end{document}